\documentclass[twocolumn,showpacs,preprintnumbers,amsmath,amssymb,floatfix,prc]{revtex4}
\usepackage{graphicx}
\usepackage{dcolumn}
\usepackage{bm}

\def\beq{\begin{equation}}
\def\eeq{\end{equation}}
\def\be{\begin{eqnarray}}
\def\ee{\end{eqnarray}}

\newcommand{\lsim}{
 \mathrel{\setbox0=\hbox{$<$}\raise0.6ex\copy0\kern-\wd0
 \lower0.65ex\hbox{$\sim$}}}

\newcommand{\gsim}{
 \mathrel{\setbox0=\hbox{$>$}\raise0.6ex\copy0\kern-\wd0
 \lower0.65ex\hbox{$\sim$}}}

\begin{document}
\title{Gluon distributions in nuclei probed at the CERN Large Hadron Collider}
\author{Adeola Adeluyi}
\author{Carlos Bertulani}
\affiliation{Department of Physics \& Astronomy,
Texas A\&M University-Commerce, Commerce, TX 75428, USA}

\date{\today}
\begin{abstract}
Using updated gluon distributions from global fits to data, we
investigate the sensitivity of direct photoproduction of heavy quarks
and exclusive production of vector mesons to varying strength of
gluon modifications. Implications of using these processes for 
constraining nuclear gluon distributions are discussed. 
\end{abstract}
\maketitle
\vspace{1cm}
%
%

Ultraperipheral relativistic heavy ion collisions can explore several aspects of particle and nuclear physics and have been extensively discussed in the literature (for a small sample of references, see e.g. \cite{Jackson, Bertulani:1988, Cahn:1990jk,Baur:1990fx,KN99,Bertulani:1999cq,Goncalves:2001vs,KNV02,Goncalves:2003is,Bertulani:2005ru,Baltz:2007kq,AyalaFilho:2008zr}). In this article, we
investigate the sensitivity of direct photoproduction of heavy quarks
and exclusive production of vector mesons to different
gluon distribution functions. This idea, originally proposed in Ref. \cite{Goncalves:2001vs}, can be used to constrain the possible distribution functions from data on production of heavy quarks and of vector mesons. Here we report on the study of certain distribution functions, not considered so far. 

A key ingredient of our calculations, the photon flux in ultraperipheral collisions, can be evaluated
by the equivalent photon (Weizs\"acker-Williams) method \cite{Jackson}.  Improvements of the method has been documented in several publications
\cite{Bertulani:1988, Cahn:1990jk,Baur:1990fx}. For a given impact parameter ${\bf b}$, the flux of virtual photons with photon energy $k$ is ${d^3N_\gamma(k,{\bf b}) / dkd^2b} $, depending strongly on the Lorentz factor $\gamma$.
At the Large Hadron Collider (LHC) at CERN the Lorentz $\gamma_L$ factor in the laboratory frame is $7455$ for p-p and $2930$
for Pb-Pb collisions.
The photon flux also depends strongly on the  adiabaticity parameter 
 $\zeta=kb/\gamma$ \cite{Bertulani:1988, Cahn:1990jk,Baur:1990fx}:
\begin{equation}
\frac{d^3N_\gamma(k,{\bf b})}{dkd^2b} = 
\frac{Z^2\alpha \zeta^2}{\pi^2kb^2} \left[ K_1^2(\zeta) + \frac{1}{
\gamma_L^2} K_0^2(\zeta) \right] \, \, .
\label{dpf}
\end{equation}
Due to the modified Bessel functions $K_0(\zeta)$ and $K_1(\zeta)$,
the photon flux possesses the asymptotic property of an exponential
drop-off at $\zeta >1$,  above a cutoff energy determined  essentially
by the size of the nucleus, $E_{cutoff} \sim \gamma$ MeV $/b$ (fm).
The relationship between the Lorentz contraction factor associated
with the relative velocity between the colliding nuclei, and the
collider energy per nucleon, $E/A$, in GeV, is given by $\gamma
=2\gamma_L^2-1 \approx 2(1.0735E/A)^2$.

Integrating ${d^3N_\gamma(k,{\bf b}) / dkd^2b} $ over impact parameters with the constraint of no hadronic interactions yields the total 
photon flux ${dN_\gamma(k) / dk} $. An analytic expression for this
total flux, strongly dependent on the reduced adiabaticity parameter 
 $\zeta_R^{AA}=2kR_A/\gamma$  for AA collisions, or $\zeta_R^{pA} =
 k(R_p + R_A)/\gamma$ for pA collisions, is derivable in the approximation whereby Eq.~(\ref{dpf}) is integrated
over impact parameters larger than the sum of the radii of the
participants. While this is a good approximation, a relatively 
better estimate of the total flux is obtained by taking the average over the
target surface \cite{KN99,Baltz:2007kq}
\begin{eqnarray}
\frac{dN_\gamma(k)}{dk}& =&  
2 \pi \int_{2R_A}^{\infty} db \, b 
\int_0^R  {dr \, r \over \pi R_A^2} 
\int_0^{2\pi} d\phi  \nonumber \\
&\times& {d^3N_\gamma(k,b+r\cos \phi)\over dkd^2b} \, \, .
\label{npf} 
\end{eqnarray}  
Although both the analytic expression and Eq.~(\ref{npf}), evaluated
numerically, have been utilized in the present study, we report only
results using the numerical flux. The differences between the 
results from analytic and numerical fluxes are generally of the order of $10
- 15\%$.
With the knowledge of the photon flux, 
any generic total photoproduction 
cross section can be factorized into
the product of a photonuclear cross section $\sigma_{X}^{\gamma}(k)$ 
and the photon flux, $dN_\gamma/dk$, $
\sigma_{X}=\int dk ({dN_\gamma}/{dk}) \sigma_{X}^{\gamma}(k) $. 

The photonuclear processes described in the present work are dependent on 
gluon distributions in nuclei. This dependence influences the
structural characteristics of these processes, especially in the case
of the exclusive photoproduction of vector mesons, in which the gluon
distribution enters quadratically. It is a rather well-known
fact that the distributions of partons (i.e. quarks and gluons) in 
nuclei are quite different from the distributions in free nucleons,
due to the complex, many-body effects in the nuclear medium. This is
expected to manifest in experimental observables such as
the cross section and rapidity distributions.

Nuclear parton distribution functions, $F_{a}^A({\bf r},x,Q^2)$, are often 
for technical convenience expressed as a convolution of "nuclear modifications" $R_{a}^A({\bf r},x,Q^2)$ and free nucleon
parton distribution functions $f_{a}(x,Q^2)$. Here the subscript $a$ denotes a
parton species and the superscript  $A$ a particular nucleus. The
variables are the position vector ${\bf r}$, parton momentum fraction 
$x$ (Bjorken-$x$), and a hard scale (factorization scale) $Q^2$.
However, since limited availability of data does not permit a
determination of the spatial dependence, current nuclear parton
distributions from global fits are functions of $x$ and $Q^2$ only. 
The nuclear effects encoded in the nuclear modifications 
$R_{a}^A(x,Q^2)$ can be categorized based on different intervals in $x$.
At small values of $x$ ($x \lesssim 0.04$), we have the phenomenon
generally referred to as shadowing. This is a depletion, in the sense
that in this interval, the nuclear parton distributions are smaller compared
to the corresponding distributions in free nucleons, i.e. $R_a^A < 1$. 
Antishadowing, which is an
enhancement ($R_a^A > 1$), occurs in the range $0.04 \lesssim x \lesssim
0.3$. Another depletion, the classic EMC effect \cite{Aubert:1983xm}, is
present in the interval $0.3 \lesssim x \lesssim 0.8$, while for 
$x > 0.8$, the Fermi motion region, we have another enhancement. 
It is important to note that although both shadowing and
the EMC effect (antishadowing and Fermi motion) correspond to
depletion (enhancement), the physical principles and mechanisms 
governing these phenomena are quite different. Further details can be
found in \cite{Geesaman:1995yd,Piller:1999wx,Armesto:2006ph,Kolhinen:2005az} 

The determination of gluon distributions in both nucleons and nuclei
is a highly nontrivial task. In the usual determination of parton
distributions from global fits to data, a preponderance of the 
experimental data is from Deeply Inelastic Scattering (DIS)
and Drell-Yan (DY)processes. 
Since gluons are electrically neutral, their distributions
cannot be directly extracted from DIS; they are inferred from sum
rules and the $Q^2$ evolution of sea quarks distributions. 
The situation is even worse in the nuclear case: the
available data is much less than for nucleons, and there is the added
complication of a mass dependence. It is therefore not unusual for
nuclear gluon distributions from different global fits to differ 
significantly, especially in the magnitude of the various nuclear
effects (shadowing, antishadowing, etc). This is especially obvious
at low $Q^2$ (i.e. around their initial starting scales) since 
evolution to high $Q^2$ tends to lessen the differences.  
Earlier global analyses 
\cite{Eskola:1998df,deFlorian:2003qf,Shad_HKN,Hirai:2007sx} relied heavily on
fixed-target nuclear deep-inelastic scattering
(DIS) and Drell-Yan (DY) lepton-pair production, with the attendant
low precision and weak constraints on nuclear gluon
distributions. The use of extended data set, incorporating data on
inclusive hadron production in deuteron-gold collisions, has been
pioneered in \cite{Eskola:2008ca,Eskola:2009uj}, 
with better constraints on gluon modifications. We should also mention 
the approach in \cite{Frankfurt:2003zd} which utilizes the Gribov picture of 
shadowing. Despite all these advances the nuclear gluon
distribution is still the least constrained aspect of global fits to 
nuclear parton distributions, as large uncertainties still persist at 
both small and large $x$.
    
Four recent gluon distributions are utilized in the present study. For
the nucleon gluon distributions we use the Martin-Stirling-Thorne-Watts
(MSTW08) parton distributions \cite{Martin:2009iq}.
In the nuclear case we use three nuclear modification sets.
Two sets are by  Eskola, Paukunnen, and Salgado, namely EPS08 and EPS09 
\cite{Eskola:2008ca,Eskola:2009uj}. 
The third is the Hirai-Kumano-Nagai (HKN07) distributions 
\cite{Hirai:2007sx}. The gluon distributions from MSTW08 serve two purposes: 
as the free nucleon distributions used in conjunction with
nuclear modifications, and also as a "special" nuclear gluon
distribution in the absence of nuclear effects. The latter usage is
particularly useful for highlighting the influence of the various
nuclear effects on observables. Our calculations are to leading order
(LO); thus all distributions are evaluated at this order.   

\begin{figure}[!htb]
\begin{center} 
\includegraphics[width=8.5cm, height=8.5cm, angle=270]{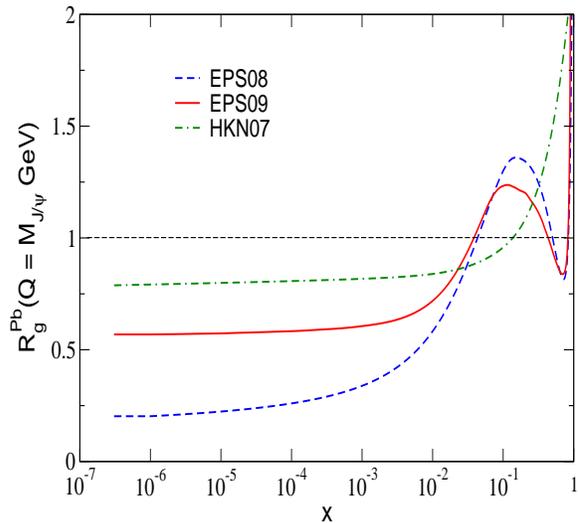}
\end{center}
\caption[...]{(Color Online) Nuclear gluon modifications in Pb,
 $R_{g}^{Pb}(x,Q^2 = M_{J/\Psi}^2)$, from EPS08 (dashed line), 
 EPS09 (solid line), and HKN07 (dash-dotted line) respectively.}
\label{fig:RgPb_Mjpsi}
\end{figure}

These four gluon distributions have different characteristics due to
the different strength of their nuclear modifications. As already
stated, one can view the MSTW08 gluon distributions as nuclear gluon
distributions in the limit of zero nuclear effects ( $R_a^A = 1$).
In Fig.~\ref{fig:RgPb_Mjpsi} we show the nuclear modifications from 
EPS08, EPS09, and HKN07 at the factorization scale $Q^2 =
M_{J/\Psi}^2$, appropriate for the elastic photoproduction of the $J/\Psi$ 
meson.
At this scale, HKN07 has a rather weak gluon shadowing which extends
well into the antishadowing region, no antishadowing and  
gluon EMC effect, and an early onset of Fermi motion. 
EPS09 exhibits a moderately strong shadowing, and appreciable
antishadowing and EMC effect, with a quite strong Fermi motion. 
Nuclear modifications are strongest in EPS08: an especially strong 
shadowing, and substantial antishadowing, EMC, and Fermi motion. 
Thus in terms of shadowing we have 
a progression from zero effects to weak effects, moderate
(intermediate) effects, then to strong effects as one progresses from 
MSTW08 to EPS08. We have not included uncertainties from gluon 
distributions in this study. Further discussions on uncertainties in
nuclear parton distributions can be found in \cite{Eskola:2009uj}.       

From the point of view of a Fock space decomposition, photon
interactions with hadrons and nuclei can be direct or resolved.
In direct interactions the photon behave as a pointlike
particle ("bare photon") while in resolved interactions the 
photon fluctuates into a quark-antiquark state or an even more complex 
partonic state. In the present study we will focus attention only 
on the direct photon contribution. The contribution of resolved photon 
processes in ultraperipheral heavy
ion collisions is treated in detail in Ref. \cite{KNV02}.  

Photon-gluon fusion leading to the production of a heavy quark pair
is the dominant subprocess when high energy photons are incident on a 
nucleus. Due to the high energies
involved, perturbative QCD is applicable, and the cross section 
can be expressed as a convolution of the partonic cross section for the
subprocess $\gamma g \rightarrow q \overline{q}$ and the
nuclear gluon distribution:
\begin{eqnarray}
\sigma^{\gamma A\rightarrow q\overline{q}X}\, (W_{\gamma A}) &=&
\int_{x_{min}}^{1} dx
 \,  \sigma^{\gamma g \rightarrow q\overline{q}}(W_{\gamma g}) \, G_A(x,Q^2)\,\,,
\label{ppcs}
\end{eqnarray}
with $x$ the momentum fraction carried by the gluon,
and $x_{min}=4m_q^2/W_{\gamma A}^2$. Here $m_q$ is the mass of the heavy
quark (charm or bottom), $W_{\gamma A}$ ($W_{\gamma g}$) denotes 
the center-of-mass energy of the photon-nucleus (photon-gluon) system, 
and $G_A(x, Q^2)$ is the nuclear gluon distribution.
The pQCD factorization scale, $Q$, is quite arbitrary; the cross
section is thus scale dependent. In the present study we use two
different scales in order to assess the magnitude of scale dependency: 
a dynamic scale $Q^2=W_{\gamma g}^2=\hat{s}$ as in \cite{Gluck:1978bf}, and 
a static scale $Q^2 = 4m_c^2 (m_b^2)$ for charm (bottom) respectively, as
in \cite{KNV02}. We take $m_c = 1.4$ GeV and $m_b = 4.75$ GeV for
consistency with the MSTW08 parton distributions. 

The cross section for the photon-gluon fusion leading to quark pair
production is \cite{Gluck:1978bf,JonesWyld,FriStreng78}:
\begin{eqnarray}
&& \sigma^{\gamma g \rightarrow q \overline{q}} \, (W_{\gamma g})= \frac {2\pi\,\alpha_{em}\,\alpha_s(Q^2)\,e_q^2}
{W_{\gamma g}^2} \nonumber \\ 
&\times&\left[(1+\beta -\frac{1}{2}\beta^2) 
\ln(\frac{1+\sqrt{1-\beta}}
{1-\sqrt{1-\beta}})  -(1+\beta) \sqrt{1-\beta}\right] \,\,,
\label{plcs}
\end{eqnarray}
with $e_q$ the electric charge of the quark, $\alpha_{em}$ the 
electromagnetic coupling constant, and
$\beta=4\,m_q^2/W_{\gamma g}^2$.
The strong coupling constant,
$\alpha_{s}(Q^{2})$, needed for the calculation, is evaluated to one
loop at the scale $Q^{2}$
using the evolution code contained in the MSTW08 package.
The total photoproduction cross section $\sigma( A[\gamma]A
\rightarrow A q \overline{q} X)$ is obtained by convoluting 
the equivalent photon flux, $dN_\gamma(k)/dk$ with 
$\sigma^{\gamma A\rightarrow q\overline{q}X}\, (k)$:
\begin{equation}
\sigma( A[\gamma]A \rightarrow A q \overline{q} X) = \int dk 
\, \frac{dN_\gamma(k)}{dk} \, \sigma^{\gamma A
  \rightarrow q \overline{q}X}(k) \, .
\label{tppcs}%
\end{equation}

The final state $q\overline q$ rapidity is dependent
on the photon energy $k$ and the gluon momentum fraction $x$: 
$x = (W_{\gamma g}/W_{\gamma A}) e^{y}$. Thus changing variable from
$k$ to $y$ and differentiating ($d\sigma/dy = kd\sigma/dk$), the 
differential cross section with respect to rapidity is thus
$
{d \sigma^{\gamma A \rightarrow q \overline{q}X}}/{dy} = (k {dN_\gamma(k)}/{dk}) \sigma^{\gamma A
\rightarrow q \overline{q}X}(k)$.

Both the total cross section and the rapidity distribution involve the
product of the photon flux and the photonuclear cross section. The flux decreases exponentially
with increasing photon energy $k$ while the cross section, due to its
dependence on $x_{min}$, increases with $k$ since $x_{min}$ is
inversely related to $k$. This interplay not only decisively
influences the  magnitude of the total cross section and rapidity
distributions, but also the relative contributions of the various
$x$-dependent nuclear modifications of the gluon distribution in 
$G_A(x,Q^2)$. For instance, gluon shadowing is of less importance in  
photoproduction of bottom quarks compared to charm quarks. This is due 
to the the fact that the advent of shadowing contribution  
in $b\bar{b}$ production occurs at a larger
photon energy where the flux is more suppressed than in  $c\bar{c}$ 
photoproduction where the onset occurs at a lower photon energy, and
thus with a larger usable photon flux availability. 

The differential cross section for the elastic photoproduction of a
vector meson $V$ on a nucleus $A$ in the exclusive process 
$A[\gamma] A \rightarrow AAV$ can be written as
$
{d \sigma^{\gamma A \rightarrow VA}}/{dt} = \left.  {d \sigma^{\gamma A \rightarrow VA}}/{dt} \right| _{t=0} | F(t) |^{2}
$,
where $d \sigma^{\gamma A \rightarrow VA}/ dt|_{t=0}$ is the forward
scattering amplitude and
$F(t)$ is the nuclear form factor. The dynamical
information is encoded in the forward scattering amplitude while
the momentum transfer of the elastic scattering is determined by the
form factor, which is dependent on the spatial attributes of the target
nucleus. 

Diverse mechanisms have been employed in the evaluation of the
dynamical content of the forward scattering amplitude for heavy
mesons. In this study we use the simple amplitude calculated from leading order 
two-gluon exchange in perturbative QCD \cite{Ryskin,Brodsky:1994kf}
and corrected for other 
relevant effects (such as relativistic corrections, inclusion of the
real part of the scattering amplitude, next-to-leading order NLO
effects, etc, see for instance \cite{Ryskin:1995hz,Frankfurt:1997fj}) through a 
phenomenological multiplicative correction factor
$\zeta_V$. 
\begin{figure}[!htb]
\begin{center} 
\includegraphics[width=8.5cm, height=8.5cm, angle=270]{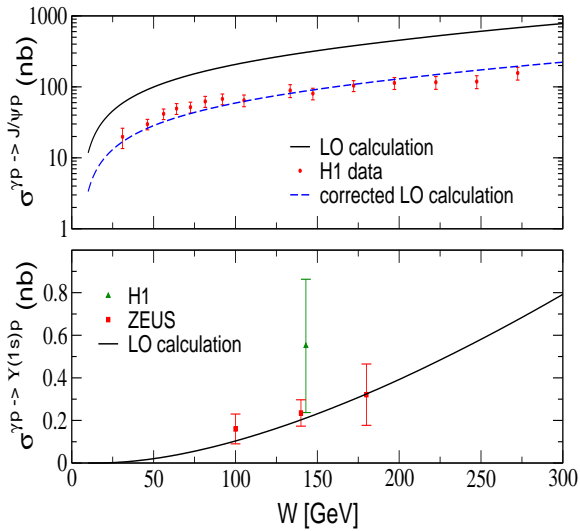}
\end{center}
\caption[...]{(Color Online) Cross section for photoproduction 
of $J/\Psi$ ($\Upsilon$) as a function of energy $W_{\gamma
  p}$. In the upper panel the solid line is the LO result while 
the dashed line is the corrected LO result using $\zeta_{J/\Psi} =
1/3.5$, with data is from \cite{Adloff:2000vm}. In the lower panel data 
taken from \cite{Adloff:2000vm,Chekanov:2009zz} and solid line depicts 
LO result.}
\label{fig:hera_jpsi_upsi}
\end{figure}
For elastic photoproduction on protons, the corrected LO 
scattering amplitude can be written as:
\begin{equation}
\left.  \frac{d \sigma^{\gamma p \rightarrow Vp}}{dt} \right| _{t=0} =
\zeta_V \frac{16\pi^3 \alpha_{s} ^{2} \, \Gamma_{ee}}{3 \alpha M_{V} ^{5}})
\left[  x g_p(x,Q^{2}) \right] ^{2} .
\label{epvm_p}
\end{equation}
Here, $M_V$ is the mass of the vector meson ($J/\Psi$ and
$\Upsilon(1s)$ in the present study), $x = M_V^2/W_{\gamma p}^2$ is the fraction of the nucleon momentum carried by the
gluons, and $g_p(x,Q^{2})$ is the gluon distribution in a proton,
evaluated at a momentum transfer $Q^{2} = (M_{V}/2)^{2}$.
Eq.~\ref{epvm_p} is easily generalized to the nuclear case:
\begin{equation}
\left.  \frac{d \sigma^{\gamma A \rightarrow VA}}{dt} \right| _{t=0} =
\zeta_V \frac{16\pi^3 \alpha_{s} ^{2} \, \Gamma_{ee}}{3 \alpha M_{V} ^{5}})
\left[  x G_A(x,Q^{2}) \right] ^{2} ,
\label{epvm_A}
\end{equation}
where $G_A(x,Q^{2}) = g_p(x,Q^{2}) \times R_g^A(x,Q^{2})$ is the
nuclear gluon distribution and $ R_g^A(x,Q^{2})$ the gluon modification.

The correction factor $\zeta_V$ is estimated by constraining the 
calculated cross sections for elastic vector mesons photoproduction on protons, 
$\sigma^{\gamma p \rightarrow Vp}(W_{\gamma p})$, to reasonably
reproduce the photoproduction data from HERA: \cite{Adloff:2000vm} 
for $J/\Psi$ and
\cite{Adloff:2000vm,Breitweg:1998ki,Chekanov:2009zz} for $\Upsilon (1s)$. 
$\sigma^{\gamma p \rightarrow Vp}$ is obtained through 
$\sigma^{\gamma p \rightarrow Vp} = (1/b) \left.  {d \sigma^{\gamma p \rightarrow Vp}}/{dt} \right| _{t=0}$ 
with slope parameter $b$. Using $b = 4.5$ GeV$^{-2}$, we have 
$\zeta_{J/\Psi} = 1/3.5$ and $\zeta_{\Upsilon (1s)} = 1.0$, and the
results are displayed in Fig.~\ref{fig:hera_jpsi_upsi}.  
$\zeta_{\Upsilon (1s)} = 1.0$ for the elastic photoproduction of
$\Upsilon$ underestimates older HERA data
\cite{Adloff:2000vm,Breitweg:1998ki} 
but seems to be adequate for the newer analysis in
\cite{Chekanov:2009zz}. 
This could indicate the smallness of higher-order effects 
and other corrections, whereas they are of major importance in 
$J/\Psi$ production.

\begin{figure}[!htb]
\begin{center} 
\includegraphics[width=8.5cm, height=8.5cm, angle=270]{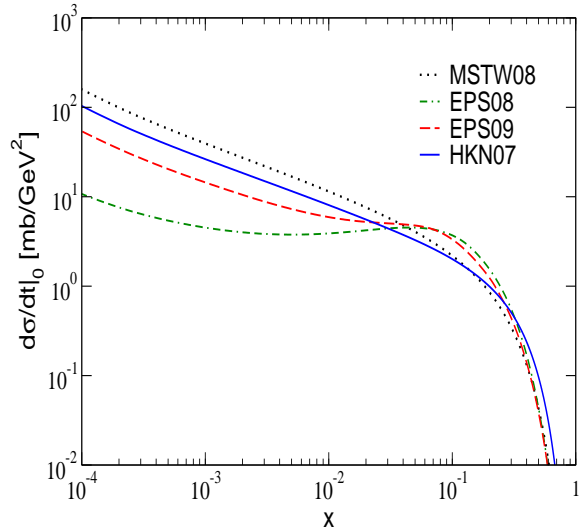}
\end{center}
\caption[...]{(Color Online) Forward scattering amplitude for elastic
  photoproduction of $J/\Psi$ in PbPb collisions as a function of
  momentum fraction $x$ for four different gluon distributions. Dotted
  line depicts result using the MSTW08 gluon distribution (no nuclear
  modifications). Dashed, dot-dashed, and solid lines are results
  from nuclear-modified gluon distributions from EPS09, EPS08, and
  HKN07 parton distributions respectively.}
\label{fig:dsdt0xjsi}
\end{figure}

In contrast to photoproduction of heavy quarks, the quadratic dependence of the
differential cross section on the gluon distribution
has the significant implication of making exclusive vector
meson production a very sensitive probe of nuclear gluon
modifications. This is apparent from Fig.~\ref{fig:dsdt0xjsi} where
the forward scattering amplitude for $J/\Psi$ production in PbPb
collisions at LHC energy is plotted as a function of $x$ for the
four gluon distributions under consideration. The different
characteristics displayed in  Fig.~\ref{fig:RgPb_Mjpsi} are clearly 
manifested.

The nuclear form factor, $F(t)$, is given by the Fourier transform of
the nuclear density distribution:
$F(t) = \int d^3r \, \rho(r) \, e^{i{\bf q} \cdot {\bf r}}$,
where $q$ is the momentum transferred. 
For a heavy nucleus it is customary to model the density
distribution as a Woods-Saxon distribution with parameters from 
electron scattering, $\rho(r)= \rho_0 /[1 + e^{[(r-R_{A})/d]}]$,
with central density $\rho_0$, radius $R_{A}$ and skin depth $d$.
For $^{208}$Pb in use at the LHC,
$\rho_0 = 0.16$/fm$^{3}$, $R_{A} = 1.2 A^{1/3}$ fm, and $d =
0.549$ fm \cite{DeJager:1974dg}. 
Since the Fourier transform of a Woods-Saxon density distribution
does not admit of an analytic form, we employed the 
commonly-used modified
hard sphere (a convolution of a hard sphere with a Yukawa term)
\cite{KN99,DN,Bertulani:1999cq} to approximate 
$\rho(r)$ in $F(t)$: 
\begin{eqnarray}
F(q=\sqrt{|t|}) &=& {4\pi\rho_0\over Aq^3}
\bigg[\sin(qR_A)-qR_A\cos(qR_A)\bigg] \nonumber \\
 &\times& \bigg[{1\over1+a^2q^2}\bigg]
\, .
\label{hsyff}
\end{eqnarray} 
The range of the Yukawa term, $a$, is $0.7$ fm, and the form factor is
a simple product of the Fourier transforms of the hard sphere and the
Yukawa term. 

The photonuclear cross section is thus given by
\begin{equation}
\sigma^{\gamma A\rightarrow VA}(k) = \frac{d\sigma^{\gamma A\rightarrow
VA}}{dt}\bigg|_{t=0} \int_{t_{min}(k)}^\infty dt |F(t)|^2
\end{equation}
Here $t_{min}(k)=(M_v^2/4k\gamma_{L})^2$, as is appropriate for narrow 
resonances \cite{klein_prl}.
The total cross section is a convolution of
the photonuclear cross section and the photon flux:
\begin{eqnarray}
\sigma^{A[\gamma]A\rightarrow AAV} = \int dk \frac{dN_\gamma(k)}{dk}
\sigma^{\gamma A\rightarrow VA}(k) = \nonumber \\
\int dk \frac{dN_\gamma(k)}{dk} \int_{t_{min}(k)}^\infty
dt \frac{d\sigma^{\gamma A\rightarrow VA}}{dt}\bigg|_{t=0} 
|F(t)|^2 \, .
\end{eqnarray}
 
It is often of practical interest to represent the cross section in
terms of the rapidity of the vector meson. The photon energy, $k$, is
related to the rapidity, $y$, by $k= (M_{V}/2) \exp(y)$. 
Using this relationship, the differential
cross section with respect to rapidity is given by
$
{d \sigma^{\gamma A \rightarrow VA}}/{dy} =( k {dN_\gamma(k)}/{dk}) \sigma^{\gamma A
\rightarrow VA}(k)$.
Thus with a knowledge of the photon
flux the differential cross section, $d \sigma/dy$, is a direct
measure of the vector meson photoproduction cross section for a given photon
energy.

We now discuss the results of our calculations for both the inclusive 
photoproduction of heavy quarks ($c\bar{c}$ and $b\bar{b}$) and the 
exclusive production of vector mesons ($J/\Psi$ and $\Upsilon (1s)$). 
For the cross sections for heavy quarks we present results 
for the two scales employed: a dynamic scale $Q^2=\hat{s}$ and a 
static scale $Q^2=4m_c^2$ ($Q^2=m_b^2$) for charm (bottom), while 
for the vector mesons we present both the LO and corrected 
LO results. In the case of rapidity distributions we show only the 
result for the dynamic scale for heavy quarks, and the corrected LO 
for the vector mesons.
Since we deal with symmetric PbPb collisions, both 
nuclei can serve as source/target and the total rapidity
distribution is the sum of both, and symmetric about $y=0$. 
\begin{figure}[!htb]
\begin{center} 
\includegraphics[width=8.5cm, height=8.5cm, angle=270]{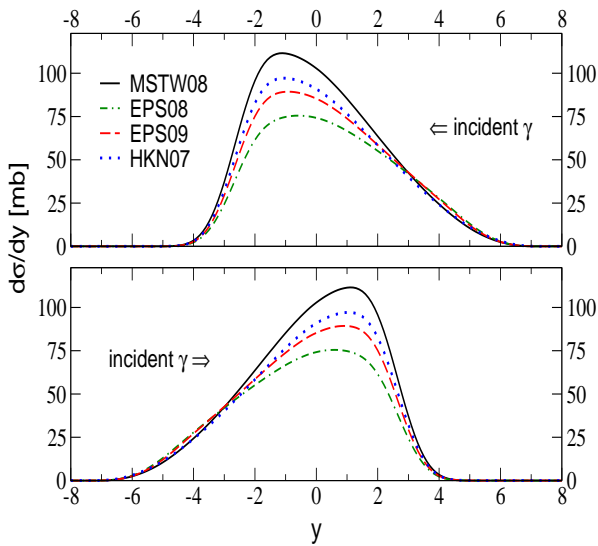}
\end{center}
\caption[...]{(Color Online) Rapidity distributions of $c\bar{c}$ 
  photoproduction in PbPb collisions at the LHC 
  for photons incident from the right (upper panel) and from the 
  left (lower panel). 
  Solid line depicts result using the MSTW08 gluon distribution (no nuclear
  modifications). Dashed, dot-dashed, and dotted lines are results
  from nuclear-modified gluon distributions from EPS09, EPS08, and
  HKN07 parton distributions respectively.}
\label{fig:hqrapccnypy}
\end{figure}

\begin{figure}[!htb]
\begin{center} 
\includegraphics[width=8.5cm, height=8.5cm, angle=270]{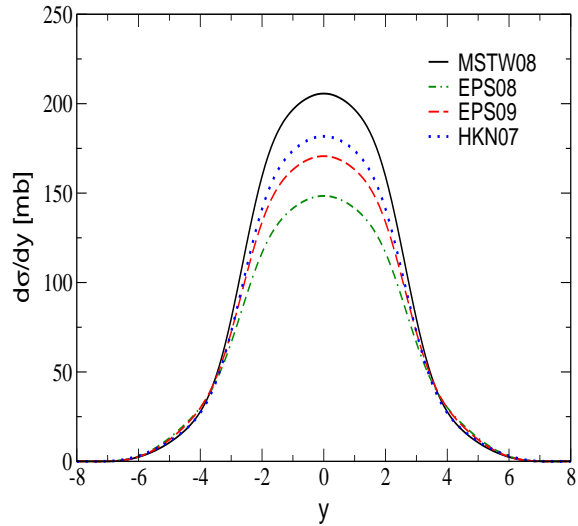}
\end{center}
\caption[...]{(Color Online) Total rapidity distributions of the
  photoproduction of $c\bar{c}$ in PbPb collisions at the LHC. Solid
  line depicts result using the MSTW08 gluon distribution (no nuclear
  modifications). Dashed, dot-dashed, and dotted lines are results
  from nuclear-modified gluon distributions from EPS09, EPS08, and
  HKN07 parton distributions respectively.}
\label{fig:hqrapcc}
\end{figure}

The sensitivity of heavy quark photoproduction to nuclear gluon modifications
is more transparent in rapidity distributions than in total cross
sections. In Fig.~\ref{fig:hqrapccnypy} we show the rapidity distributions
for $c\bar{c}$ production in ultraperipheral PbPb collisions at the
LHC, employing the four gluon distributions described earlier. The
upper panel depicts the distributions with the incident photon 
coming from the right, while the lower panel shows the distributions 
with the incident photons from the left. The total, which is the sum 
of both panels, is displayed in Fig.~\ref{fig:hqrapcc}.

As mentioned earlier, the value of $x_{min}(k)$ determines
the nuclear effects contributing to  the photonuclear cross section 
$\sigma^{\gamma A\rightarrow  q\overline{q}X}\, (k)$. Since the
rapidity distribution $d\sigma/dy$ is directly proportional to this
cross section, it is rather straightforward to elucidate the features
of the rapidity distribution based on the relative contributions of
the relevant nuclear effects. Fig.~\ref{fig:RgPb_Mjpsi} is thus quite 
helpful in understanding the characteristics of the distributions displayed 
in Fig.~\ref{fig:hqrapccnypy} and, by extension, Fig.~\ref{fig:hqrapcc}.

Let us start from midrapidity ($y = 0$) and move towards higher positive
values of $y$. In the upper panel of Fig.~\ref{fig:hqrapccnypy} 
(photons from right) the contribution from shadowing becomes progressively 
smaller, such that at around $y = 3$ shadowing is negligible and 
antishadowing becomes the dominant contributor. At around $y = 6$ 
antishadowing peters out, and the distributions are subject only 
to EMC/Fermi motion effects. On the other hand, for the lower 
panel in Fig.~\ref{fig:hqrapccnypy} (photons from left), shadowing 
becomes progressively stronger and more dominant for increasing $y$. 
The effect of shadowing in this region is mitigated by the rather 
strong suppression of the photon flux at large $k$; the distributions 
thus rapidly die out with increasing positive $y$.
The opposite trend is observed for increasingly negative $y$ values 
starting from midrapidity. Here, for photons incident from the right
(upper panel), increasing negative values of $y$ implies increasing
shadowing and flux suppression. For photons incident from the left
(lower panel), the transition is from shadowing to EMC/Fermi motion
effects.    

We now consider the characteristics of the total rapidity distributions due 
to the amalgamation of the contrasting tendencies exhibited in the   
upper and lower panels of Fig.~\ref{fig:hqrapccnypy}.
Shadowing is the dominant nuclear effect for $-3 < y < 3$, and the rapidity
distributions in this region reproduce the observed trend of 
gluon shadowing strength as exhibited in  Fig.~\ref{fig:RgPb_Mjpsi}.
MSTW08 with its zero gluon shadowing gives the largest rapidity
distribution while EPS08, with its strong gluon shadowing, gives the
smallest. Due to strong flux suppression, shadowing is most markedly 
apparent for the rapidity range $-2 \lesssim y \lesssim 2$. 
This range therefore provides a good window to discriminate among 
different gluon shadowing scenarios.

The rapidity intervals  $3 < y < 6$ 
corresponds to $x_{min}$ in the antishadowing region (deep shadowing) 
for right (left) incident photons and vice versa for $-6 < y < -3$. 
Due to the photon flux suppression 
in the deep shadowing region, the rapidity distributions are 
sensitive mainly to  antishadowing in addition to both EMC effect and 
Fermi motion. Since both EPS08 and EPS09 have substantial
antishadowing, their rapidity distributions reflect this, 
being slightly larger than those from HKN07 and MSTW08. The
discriminatory power here is not as appreciable as in the shadowing
case though, due largely to the smallness of the distributions.

For both rapidity ranges $y < -6$ and $y > 6$, $x_{min} > 0.2$ and the 
relevant contributing nuclear effects are the EMC and Fermi motion
since the contribution from antishadowing is small, and that from
shadowing practically nonexistent by virtue of flux suppression. Due to the
behavior of HKN07 in this interval (no EMC effect, only enhancement), 
$d\sigma/dy$ from HKN07 is largest. Both EPS08 and EPS09 nuclear
modifications exhibit EMC effect and Fermi motion, and the destructive 
interference from both effects render their rapidity distributions to
practically coincide with that from MSTW08.

\begin{table}
\caption{Total cross sections for direct photoproduction of 
$c\bar{c}$ in ultraperipheral PbPb collisions at the LHC. All
cross sections are in millibarns (mb).}
\begin{tabular}[c]{lccc c c}
\hline
Gluon Distribution  && $Q^2 = \hat{s}$ && $Q^2 = 4m^2_c$  \\
\hline
MSTW08              && 1170		       && 1090	\\
EPS08               && 890		       && 780	\\
EPS09	            && 1000		       && 910	      \\
HKN07               && 1080	               && 1000    \\
\end{tabular} 
\label{tcccbar}
\end{table}

In Table~\ref{tcccbar} we present the total cross section for the direct 
photoproduction of $c\bar{c}$ at two different scales as discussed 
earlier. For both scales the cross sections exhibit a clear trend in conformity 
with the relative strength of gluon shadowing. MSTW08 with no
modifications gives the largest cross section while EPS08 gives the
smallest, due to its strong gluon shadowing. It is also
noteworthy that the static scale gives lower cross sections relative
to the dynamic scale. The difference between the two scales 
increases progressively with increasing shadowing, from $\approx 7\%$ 
for MSTW08 to around $13\%$ for EPS08. Overall, the total cross
section seems also a good discriminator of different gluon shadowing scenarios.

Our cross sections can be compared with results from other studies 
on $c\bar{c}$ photoproduction. In \cite{Bertulani:2005ru,RV} the cross section
with no shadowing (equivalent to our MSTW08) is $1250$ mb, from EKS98
 \cite{Eskola:1998df} (somewhat akin to EPS09) it is $1050$ mb, 
while from FGS (strong shadowing as in EPS08) \cite{Frankfurt:2003zd} 
it is $850$ mb. Likewise in \cite{KNV02} the no shadowing
cross section is $1790$ mb while the cross section from EKS98 is 
$1500$ mb. The study in \cite{Goncalves:2003is} gives the cross section as 
$2056$ mb. Overall, our results are closest to those reported in 
\cite{Bertulani:2005ru,RV}. Differences in results are not only
attributable to the different gluon distributions and photon 
fluxes used in the earlier studies, but also possibly to different 
values of the running strong coupling, $\alpha_s(Q^2)$, which enters 
multiplicatively in the expression for the photon-gluon cross section.    

\begin{figure}[!htb]
\begin{center} 
\includegraphics[width=8.5cm, height=8.5cm, angle=270]{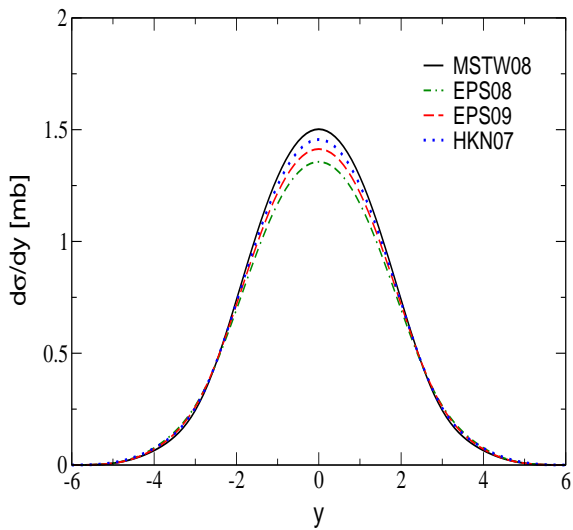}
\end{center}
\caption[...]{(Color Online) Total rapidity distributions of $b\bar{b}$ 
  photoproduction in PbPb collisions at the LHC. Solid
  line depicts result using the MSTW08 gluon distribution (no nuclear
  modifications). Dashed, dot-dashed, and dotted lines are results
  from nuclear-modified gluon distributions from EPS09, EPS08, and
  HKN07 parton distributions respectively.}
\label{fig:hqrapbb}
\end{figure}

We now discuss our result for the total cross sections and 
rapidity distributions for $b\bar{b}$ production. The cross sections are 
orders of magnitude less than in $c\bar{c}$ production, and also
correspondingly exhibit less sensitivity to nuclear modifications.  
In Fig.~\ref{fig:hqrapbb} we show the rapidity distribution for $b\bar{b}$
in ultraperipheral PbPb collisions at the LHC.  
Shadowing dominates in the rapidity interval $-2 < y < 2$ and is 
most clearly manifested in the rapidity window $-1 < y < 1$. Thus 
this interval presents the best sensitivity to shadowing effects 
in $b\bar{b}$ production. Although less marked, the progression 
of rapidity distribution with relative shadowing strength follows the 
trend observed in $c\bar{c}$ production: MSTW08 still gives the
largest distribution while EPS08 gives the smallest.

As in the case of $c\bar{c}$ production, there is a slight
manifestation of the influence of antishadowing around  
$y = -3$ and $y = 3$. The distributions practically overlap
for $y < -4$ and $y > 4$; thus overall, the interval 
$-1 \lesssim y \lesssim 1$ seems to afford the best sensitivity to 
nuclear effects, in this case primarily shadowing. More detailed 
treatment of the $x$-dependence of $b\bar{b}$ production at the 
LHC is presented in \cite{Strikman:2005yv}. 

\begin{table}[!htb]
\caption{Total cross sections for direct photoproduction of 
$b\bar{b}$ in ultraperipheral PbPb collisions at the LHC. All 
cross sections are in millibarns (mb).}
\begin{tabular}[c]{lccc c c}
\hline
Gluon Distribution  && $Q^2 = \hat{s}$ && $Q^2 = m^2_b$  \\
\hline
MSTW08              && 6.2	       && 7.0	\\
EPS08               && 5.8	       && 6.2	\\
EPS09	            && 6.0	       && 6.6	\\
HKN07               && 6.1	       && 6.7    \\
\end{tabular}
\label{tcbbbar}
\end{table} 

The total photoproduction cross sections for $b\bar{b}$ from the four 
gluon distributions are presented in  Table~\ref{tcbbbar}. 
It is readily observed from the table that nuclear modifications 
have manifestly weaker effects on $b\bar{b}$ production relative 
to $c\bar{c}$, since the results from two widely different
distributions like MSTW08 and EPS08 are almost in the same ballpark. 
This observation is also apparently scale independent. Unlike the 
$c\bar{c}$ case though, the static scale gives larger cross sections 
than the dynamic scale, and the difference decreases with relative 
shadowing strength.

Our cross sections can be compared with results from other studies 
on $b\bar{b}$ photoproduction. In \cite{Bertulani:2005ru,RV} the cross section
with no shadowing is $4.9$ mb, from EKS98 it is $4.7$ mb, while from
FGS it is $4.4$ mb. Likewise in \cite{KNV02} the no shadowing
cross section is $0.718$ mb while the cross section from EKS98 is 
$0.686$ mb. The study in \cite{Goncalves:2003is} gives the cross section as 
$20.1$ mb. Again, our results are closest to the values reported in 
\cite{Bertulani:2005ru}. The comment concerning the differencies in the
$c\bar{c}$ results is also applicable here. Additional details
can be found in \cite{Goncalves:2003is}. 

It is pertinent at this point to remark on the limitations inherent 
in our calculated cross sections for both $c\bar{c}$ and $b\bar{b}$ 
photoproduction. The results are to leading order; thus higher-order 
effects have not been taken into account either explicitly or through 
a phenomenological correction factor. In addition, we have not 
included the resolved contributions which are quite sizeable 
(see \cite{KNV02}). We thus advocate that these limitations should be
borne in mind, moreso in view of the disparities in  
cross section results from the present work and previous studies.
 
We now present our results on elastic photoproduction of the $J/\Psi$ 
and $\Upsilon(1s)$ in the framework of a leading-order two-gluon exchange 
formalism in QCD.
Apart from the quadratic dependence on the gluon distribution, two
other quantities namely the integrated nuclear form factor and the
photon flux affect the attributes of both the rapidity distribution 
and total cross section. The photon flux has support at low photon
energy $k$, which translates to large $x$, while the integrated form
factor favors large $k$, or equivalently, small $x$. 

\begin{figure}[!htb]
\begin{center} 
\includegraphics[width=8.5cm, height=8.5cm, angle=270]{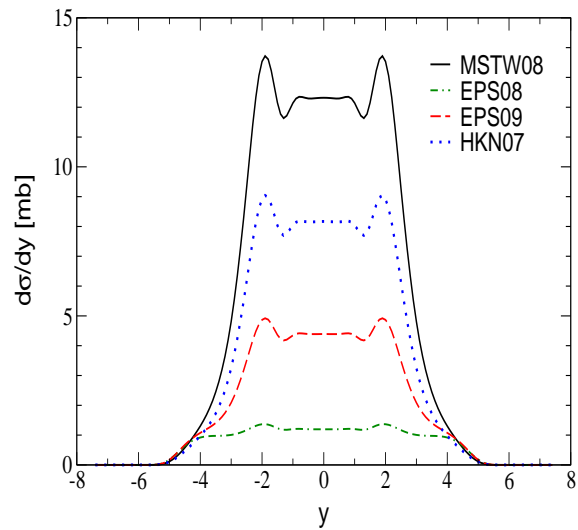}
\end{center}
\caption[...]{(Color Online) Total rapidity distributions of exclusive
  photoproduction of $J/\Psi$ in PbPb collisions at the LHC in the
  modified hard sphere density distribution approximation. Solid
  line depicts result using the MSTW08 gluon distribution (no nuclear
  modifications). Dashed, dot-dashed, and dotted lines are results
  from nuclear-modified gluon distributions from EPS09, EPS08, and
  HKN07 parton distributions respectively.}
\label{fig:jpsiyt}
\end{figure}

Fig.~\ref{fig:jpsiyt} shows the rapidity distribution of the 
$J/\Psi$ in ultraperipheral PbPb collisions at the LHC.
Shadowing is the relevant nuclear effect in the rapidity interval 
$-3 < y < 3$ and unsurprisingly, the rapidity distributions mimic 
the behavior in the shadowing region of Fig.~\ref{fig:RgPb_Mjpsi}. 
The largest rapidity distribution is given by MSTW08, followed by
HKN07, and EPS09. The smallest is by EPS08 due to its strong gluon 
shadowing. The rapidity window $-2 < y < 2$ manifestly depicts the 
significant distinction between the various gluon distributions 
arising from the quadratic dependence. Antishadowing manifests 
in the intervals $-5 < y < -4$ and $4 < y < 5$; the effect though 
is quite slight.      

\begin{table}[!htb]
\caption{Total cross sections (in mb) for elastic photoproduction of 
 $J/\Psi$ in ultraperipheral PbPb collisions at the LHC. Second 
column is the result from LO calculation while third column displays
result from corrected LO calculation.}
\begin{tabular}[c]{lccc c c}
\hline
Gluon Distribution  && LO  && Corrected LO \\
\hline
MSTW08              && 260   && 74 \\		      
EPS08               && 36   && 10 \\		       
EPS09	            && 101   && 29 \\		      
HKN07               && 173   && 49 \\	               
\end{tabular}
\label{tjpsi}
\end{table}

Table~\ref{tjpsi} shows the total cross sections for the elastic
photoproduction of the $J/\Psi$ using the four gluon distributions 
considered in our study. The leading order result is under the second
column while the corrected leading order result is under the third. 
As expected, the total cross sections reproduce the trend 
seen in the rapidity distributions: MSTW08 gives the largest cross
section while EPS08 yields the smallest. 

There have been studies of the photoproduction of $J/\Psi$ 
in ultra-peripheral collisions at LHC using 
diverse production mechanisms \cite{KN99,Goncalves:2001vs,Baltz:2007kq,
strikman_plb,per4,ivanov_kop,vicmag_prd2008,AyalaFilho:2008zr}. 
Here, since we are interested in the
sensitivity to gluon modifications, we use the simple leading order (LO) 
two-gluon exchange mechanism corrected for additional relevant 
effects through a multiplicative factor.
Our work is similar in spirit to the study reported in
\cite{AyalaFilho:2008zr} in the sense that four different gluon 
distributions were also utilized. 
Thus we compare our results to the work in 
\cite{AyalaFilho:2008zr} which reports a no shadowing 
cross section of $74$ mb, a cross section from EPS08 of $13$ mb, 
EKS98 of $39$ mb, and from DS03 \cite{deFlorian:2003qf} 
(somewhat similar to HKN07) of $61$ mb. 
While the corrected LO results are somewhat close to these values, 
the uncorrected LO results presented in Table~\ref{tjpsi} are 
consistently higher. 

In Table~\ref{tjpsi_rhic} we also present the equivalent cross sections for 
$J/\Psi$ production in ultraperipheral AuAu collisions at RHIC.
As is readily apparent, the results follow the shadowing trend 
as observed for PbPb collisions at the LHC.

\begin{table}[!htb]
\caption{Total cross sections (in $\mu$b) for elastic photoproduction of 
 $J/\Psi$ in ultraperipheral AuAu collisions at RHIC.}
\begin{tabular}[c]{lccc c c}
\hline
Gluon Distribution  && LO  && Corrected LO \\
\hline
MSTW08              && 1222   && 349 \\		      
EPS08               && 699   && 200 \\		       
EPS09	            && 868   && 248 \\		      
HKN07               && 902   && 258 \\	               
\end{tabular}
\label{tjpsi_rhic}
\end{table}

\begin{figure}[!htb]
\begin{center} 
\includegraphics[width=8.5cm, height=8.5cm, angle=270]{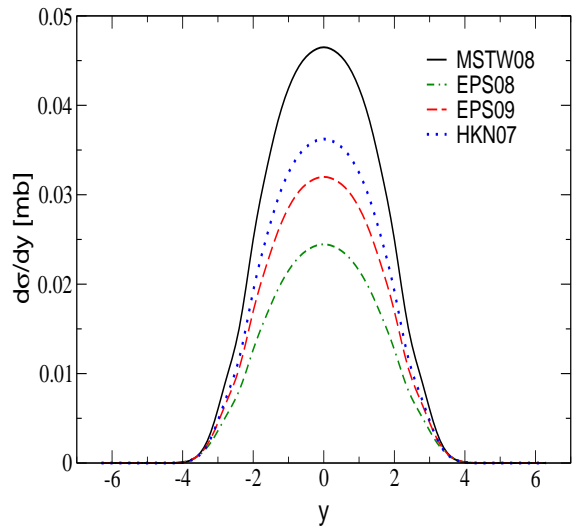}
\end{center}
\caption[...]{(Color Online) Total rapidity distributions of exclusive
  photoproduction of $\Upsilon(1s)$ in PbPb collisions at the LHC in the
  modified hard sphere density distribution approximation. Solid
  line depicts result using the MSTW08 gluon distribution (no nuclear
  modifications). Dashed, dot-dashed, and dotted lines are results
  from nuclear-modified gluon distributions from EPS09, EPS08, and
  HKN07 parton distributions respectively.}
\label{fig:upsiyt}
\end{figure}

The rapidity distributions for $\Upsilon(1s)$ production is shown in
Fig.~\ref{fig:upsiyt}, and exhibit identical trend observed   
for $J/\Psi$. Shadowing remains 
the relevant nuclear modification for practically the entire rapidity 
range shown in the figure, and is markedly manifested in the interval 
$-2 < y < 2$. Thus rapidity distribution in this interval should be 
a good discriminator of gluon shadowing strength. 

\begin{table}[!htb]
\caption{Total cross sections for elastic photoproduction of 
 $\Upsilon(1s)$ in ultraperipheral PbPb collisions at the LHC.}
\begin{tabular}[c]{lcc}
\hline
Gluon Distribution  && Cross Section ($\mu$b) \\
\hline
MSTW08              && 189  \\		      
EPS08               && 99  \\		       
EPS09	            && 130  \\		      
HKN07               && 146 \\	               
\end{tabular}
\label{tupsi}
\end{table}

In Table~\ref{tupsi} we show the total cross sections for the elastic
photoproduction of the $\Upsilon(1s)$ for the four gluon distributions 
under study. Unsurprisingly, the total cross sections reflect the progressive 
trend of the relative shadowing strength, MSTW08 giving the largest cross
section while EPS08 yields the smallest.

Previous studies of the photoproduction of $\Upsilon(1s)$ in
ultra-peripheral collisions  at LHC energies have been reported in 
\cite{strikman_jhep,klein_prl,vicmag_prd2008,AyalaFilho:2008zr}. 
As in the case of $J/\Psi$ production, we compare our results with 
the values reported in \cite{AyalaFilho:2008zr}. The no shadowing 
cross section is $163$ $\mu$b, from EPS08 $22$ $\mu$b, 
EKS98 $120$ $\mu$b, and from DS03 $148$ $\mu$b. 
Except for the case of EPS08, our results presented in
Table~\ref{tupsi} are quite close to these values. 

In conclusion, we have considered the direct photoproduction of
 heavy quarks (charm
and bottom) and elastic photoproduction of vector mesons ($J/\Psi$ 
and $\Upsilon(1s)$) in ultraperipheral PbPb collisions at LHC
energy. These two processes are dependent on nuclear gluon
distributions, and are therefore potentially useful in constraining
modifications such as shadowing and antishadowing in nuclear 
gluon distributions. In order to assess the sensitivity to 
these modifications, we have utilized four recent gluon 
distributions, chosen on the basis of the relative strength of 
their modifications. For each process we considered two observables: 
rapidity distributions and total cross sections.

In direct photoproduction of heavy quarks the gluon dependence is
linear and different modifications are superimposed due to the
integration over the momentum fraction $x$. Despite these, rapidity 
distributions for $c\bar{c}$ manifest appreciable sensitivity to
shadowing around midrapidity and a slight sensitivity to antishadowing 
at more forward and backward rapidities. Thus $c\bar{c}$
photoproduction offers good constraining potential for shadowing, and
a somewhat less potential for antishadowing. Although photoproduction of 
$b\bar{b}$ is less sensitive to modifications than $c\bar{c}$, the 
influence of shadowing is evident around midrapidity, and it thus
offers some constraining ability for shadowing. 
    
The quadratic dependence on gluon modifications makes elastic
photoproduction of vector mesons particularly attractive for 
constraining purposes. This is manifestly apparent from the rapidity 
distributions for both $J/\Psi$ and $\Upsilon(1s)$ photoproduction which
exhibit very good sensitivity to gluon shadowing over an appreciable 
range about midrapidity. Thus both offer remarkable potential in 
constraining the shadowing component of nuclear gluon distributions. 

Determination of nuclear modifications from ab-initio calculations of
cross sections is beset with difficulties. A more feasible approach 
is to compare photoproduction in proton-nucleus and nucleus-nucleus
collisions, where many theoretical uncertainties and systematic 
errors cancel (see \cite{Salgado:2011wc}). Further work along this 
line is in progress.

We acknowledge support by the US Department of Energy Grants  DE-FG02-08ER41533 and DE- FC02-07ER41457 (UNEDF, SciDAC-2) and the Research Corporation.
%


\begin{thebibliography}{99}
%
\bibitem{Jackson}
E. Fermi, Z. Physik  {\bf 29}, 315 (1924); Nuovo Cimento {\bf 2}, 143 (1925).

\bibitem{Bertulani:1988}
C.~A. Bertulani and G.~Baur, {Phys. Rep.} {\bf 163}, 299 (1988).

\bibitem{Cahn:1990jk}
R.~N. Cahn and J. D. Jackson, {Phys.\ Rev.} D {\bf 42}, 3690 (1990).

\bibitem{Baur:1990fx}
G. Baur and L. G. Ferreira Filho,
{Nucl.\ Phys.} A {\bf 518}, 786 (1990).

\bibitem {KN99}S. Klein and J. Nystrand, {Phys. Rev. C } {\bf 60}, 014903
(1999).

\bibitem{Bertulani:1999cq}
  C.~A.~Bertulani and D.~S.~Dolci,
  Nucl.\ Phys.\  A {\bf 674}, 527 (2000).
  
  
\bibitem{Goncalves:2001vs}
  V.~P.~Goncalves and C.~A.~Bertulani,
  Phys.\ Rev.\  C {\bf 65}, 054905 (2002).


\bibitem{KNV02}
S.~R.~Klein, J.~Nystrand and R.~Vogt,
Phys.\ Rev.\ C {\bf 66}, 044906 (2002).


\bibitem{Goncalves:2003is}
  V.~P.~Goncalves and M.~V.~T.~Machado,
  Eur.\ Phys.\ J.\  C {\bf 31}, 371 (2003).

\bibitem{Bertulani:2005ru}
  C.~A.~Bertulani, S.~R.~Klein and J.~Nystrand,
  Ann.\ Rev.\ Nucl.\ Part.\ Sci.\  {\bf 55}, 271 (2005).

\bibitem{Baltz:2007kq}
  A.~J.~Baltz {\it et al.},
  Phys.\ Rept.\  {\bf 458}, 1 (2008).

\bibitem{AyalaFilho:2008zr}
  A.~L.~Ayala Filho, V.~P.~Goncalves and M.~T.~Griep,
  Phys.\ Rev.\  C {\bf 78}, 044904 (2008).

\bibitem{Aubert:1983xm}
  J.~J.~Aubert {\it et al.}  [European Muon Collaboration],
  Phys.\ Lett.\  B {\bf 123}, 275 (1983).
%
\bibitem{Geesaman:1995yd}
  D.~F.~Geesaman, K.~Saito and A.~W.~Thomas,
  Ann.\ Rev.\ Nucl.\ Part.\ Sci.\  {\bf 45}, 337 (1995).
%
\bibitem{Piller:1999wx}
  G.~Piller and W.~Weise,
  Phys.\ Rept.\  {\bf 330}, 1 (2000).
%
\bibitem{Armesto:2006ph}
  N.~Armesto,
  J.\ Phys.\ G {\bf 32}, R367 (2006).

\bibitem{Kolhinen:2005az}
  V.~J.~Kolhinen,
 arXiv:hep-ph/0506287.

\bibitem{Eskola:1998df}
  K.~J.~Eskola, V.~J.~Kolhinen and C.~A.~Salgado,
  Eur.\ Phys.\ J.\ C {\bf 9}, 61 (1999).
%
\bibitem{deFlorian:2003qf}
  D.~de Florian and R.~Sassot,
  Phys.\ Rev.\  D {\bf 69}, 074028 (2004).
%
\bibitem{Shad_HKN}
  M.~Hirai, S.~Kumano and T.~H.~Nagai,
  Phys.\ Rev.\  C {\bf 70}, 044905 (2004);
  Nucl.\ Phys.\ Proc.\ Suppl.\  {\bf 139}, 21 (2005).
%
\bibitem{Hirai:2007sx}
  M.~Hirai, S.~Kumano and T.~H.~Nagai,
  Phys.\ Rev.\  C {\bf 76}, 065207 (2007).

\bibitem{Eskola:2008ca}
  K.~J.~Eskola, H.~Paukkunen and C.~A.~Salgado,
  JHEP {\bf 0807}, 102 (2008).
%
\bibitem{Eskola:2009uj}
  K.~J.~Eskola, H.~Paukkunen and C.~A.~Salgado,
  JHEP {\bf 0904}, 065 (2009).

\bibitem{Frankfurt:2003zd}
  L.~Frankfurt, V.~Guzey and M.~Strikman,
  Phys.\ Rev.\ D {\bf 71}, 054001 (2005).

\bibitem{Martin:2009iq}
  A.~D.~Martin, W.~J.~Stirling, R.~S.~Thorne and G.~Watt,
  Eur.\ Phys.\ J.\  C {\bf 63}, 189 (2009).

\bibitem{Gluck:1978bf}
  M.~Gluck and E.~Reya,
  Phys.\ Lett.\  B {\bf 79}, 453 (1978).


\bibitem {JonesWyld}\textrm{L. M. Jones and H. W. Wyld, Phys. Rev. D } {\bf 17}, 759
(1978).

\bibitem {FriStreng78}\textrm{H. Fritzsch and K. H. Streng, Phys. Lett. B}
{\bf 72}, 385 (1978).

\bibitem {Ryskin}\textrm{M. G. Ryskin, Z. Phys., C } {\bf 57}, 89
  (1993).

\bibitem{Brodsky:1994kf}
  S.~J.~Brodsky, L.~Frankfurt, J.~F.~Gunion, A.~H.~Mueller and M.~Strikman,
  Phys.\ Rev.\  D {\bf 50}, 3134 (1994)
  [arXiv:hep-ph/9402283].

\bibitem{Ryskin:1995hz}
  M.~G.~Ryskin, R.~G.~Roberts, A.~D.~Martin and E.~M.~Levin,
  Z.\ Phys.\  C {\bf 76}, 231 (1997)
  [arXiv:hep-ph/9511228].

\bibitem{Frankfurt:1997fj}
  L.~Frankfurt, W.~Koepf and M.~Strikman,
  Phys.\ Rev.\  D {\bf 57}, 512 (1998)
  [arXiv:hep-ph/9702216].

\bibitem{Adloff:2000vm}
  C.~Adloff {\it et al.}  [H1 Collaboration],
  Phys.\ Lett.\  B {\bf 483}, 23 (2000)
  [arXiv:hep-ex/0003020].

\bibitem{Breitweg:1998ki}
  J.~Breitweg {\it et al.}  [ZEUS Collaboration],
  Phys.\ Lett.\  B {\bf 437}, 432 (1998)
  [arXiv:hep-ex/9807020].

\bibitem{Chekanov:2009zz}
  S.~Chekanov {\it et al.}  [ZEUS Collaboration],
  Phys.\ Lett.\  B {\bf 680}, 4 (2009)
  [arXiv:0903.4205 [hep-ex]].


\bibitem{DeJager:1974dg}
  C.~W.~De Jager, H.~De Vries and C.~De Vries,
  Atom.\ Data Nucl.\ Data Tabl.\  {\bf 14}, 479 (1974).

\bibitem{DN} K. T. R. Davies and J. R. Nix, Phys. Rev. 
{\bf C14}, 1977 (1976).

\bibitem{Strikman:2005yv}
  M.~Strikman, R.~Vogt and S.~N.~White,
  Phys.\ Rev.\ Lett.\  {\bf 96}, 082001 (2006)
  [arXiv:hep-ph/0508296].


\bibitem{RV} R. Vogt, private communication

\bibitem{klein_prl}
S.~R.~Klein and J.~Nystrand,
Phys.  Rev. Lett.  {\bf 92}, 142003 (2004).


\bibitem{strikman_plb}
L.~Frankfurt, M.~Strikman and M.~Zhalov,
Phys.\ Lett.\ B {\bf 540}, 220 (2002).

 \bibitem{per4}
  V.~P.~Goncalves and M.~V.~T.~Machado,
  Eur.\ Phys.\ J.\  C {\bf 40}, 519 (2005).

\bibitem{vicmag_prd2008}
  V.~P.~Goncalves and M.~V.~T.~Machado,
  Phys.\ Rev.\  D {\bf 77}, 014037 (2008).

\bibitem{ivanov_kop}
  Yu.~P.~Ivanov, B.~Z.~Kopeliovich and I.~Schmidt,
  arXiv:0706.1532 [hep-ph].

\bibitem{strikman_jhep}
  L.~Frankfurt, V.~Guzey, M.~Strikman and M.~Zhalov,
  JHEP {\bf 0308}, 043 (2003).

\bibitem{vicmag_prc}
  V.~P.~Goncalves and M.~V.~T.~Machado,
  Phys. Rev. C {\bf 73}, 044902 (2006).

\bibitem{Salgado:2011wc}
  C.~A.~Salgado {\it et al.},
  arXiv:1105.3919 [hep-ph].


%
\end{thebibliography}
\end{document}